\documentclass[5p,twocolumn,number,A4]{elsarticle}

\usepackage{color} % standard color package
\usepackage{graphicx} % standard graphics package
\usepackage{latexsym,amssymb,amsmath,amsfonts} % extended symbol packages
\usepackage{mathrsfs} % mathscripts \mathscr
\usepackage{fixltx2e} % improves figure floating in Latex2e kernel

\allowdisplaybreaks[4]
\newtheorem{mylemma}{Lemma}[section]
\newtheorem{mydefinition}{Definition}[section]
\newtheorem{mytheorem}{Theorem}[section]
\newtheorem{mycorollary}{Corollary}[section]
\newtheorem{myexample}{Example}[section]

\definecolor{Light}{gray}{0.85}

\def\abs#1{\left\vert #1 \right\vert}
\def\allpoly{\mbox{$\re\langle X \rangle$}}
\def\allpolyell{\mbox{$\re^{\ell}\langle X \rangle$}}
\def\allpolyx0degn{\mbox{$P_n$}}

\def\allseries{\mbox{$\re\langle\langle X \rangle\rangle$}}

\def\allseriesell{\mbox{$\re^{\ell} \langle\langle X \rangle\rangle$}}

\def\allseriesLC{\mbox{$\re_{LC}\langle\langle X \rangle\rangle$}}

\def\allseriesmLC{\mbox{$\re^{m}_{LC}\langle\langle X \rangle\rangle$}}

\def\allseriesellLC{\mbox{$\re^{\ell}_{LC}\langle\langle X \rangle\rangle$}}

\def\allseriesXO{\mbox{$\re [[ X_0 ]]$}}

\def\bull{\rule{0.08in}{0.08in}} % square filled bullet

\newcommand{\comment}[1]{} % allows one to comment out a block of text

\def\Endallseries{{\rm End}(\allseries)}

\def\eqref#1{(\ref{#1})} % parentheses around referenced equation numbers
\def\mbf#1{\hbox{\mathversion{bold}$#1$}} % math boldface
\def\nat{{\mathbb N}} % natural numbers (AMS symbol)
\def\norm#1{\left\Vert#1\right\Vert}
\def\notin{{\not\in}}
\def\openbull{\framebox[0.08in][c]{$\;$}} % square unfilled bullet
\def\ord{{\rm ord}}

\def\re{{\mathbb R}} % real numbers (AMS symbol)

\def\shuffle{{\scriptscriptstyle \;\sqcup \hspace*{-0.05cm}\sqcup\;}}

\def\supp{{\rm supp}}
\def\begals{\[\begin{aligned}}
\def\endals{\end{aligned}\]}
\def\begce{\begin{center}}
\def\endce{\end{center}}
\def\begar{\begin{array}}
\def\endar{\end{array}}
\def\begeq{\begin{equation}}
\def\endeq{\end{equation}}
\def\begdi{\begin{displaymath}}
\def\enddi{\end{displaymath}}
\def\begdis{\begin{eqnarray*}}
\def\enddis{\end{eqnarray*}}
\def\begeqa{\begin{eqnarray}}
\def\endeqa{\end{eqnarray}}
\def\begdes{\begin{description}}
\def\enddes{\end{description}}
\def\begit{\begin{itemize}}
\def\endit{\end{itemize}}
\def\begen{\begin{enumerate}}
\def\enden{\end{enumerate}}
\def\beglar{\left[\begin{array}}
\def\endrar{\end{array}\right]}
\def\begle{\begin{mylemma}}
\def\endle{\end{mylemma}}
\def\begde{\begin{mydefinition}}
\def\endde{\end{mydefinition}}
\def\begth{\begin{mytheorem}}
\def\endth{\end{mytheorem}}
\def\begco{\begin{mycorollary}}
\def\endco{\end{mycorollary}}
\def\begprop{\begin{myproposition}}
\def\endprop{\end{myproposition}}
\def\begex{\begin{myexample}}
\def\endex{\hfill\openbull \end{myexample}}
\def\begexer{\begin{myexercise}}
\def\endexer{\end{myexercise}}

\def\begres{\noindent{\bf Remarks}:\begin{enumerate}}
\def\endres{\end{enumerate} \par}
\def\begpr{\noindent{\em Proof:}$\;\;$}
\def\endpr{\hfill\bull \vspace*{0.15in}}
\def\begtab{\begin{tabular}}
\def\endtab{\end{tabular}}
\def\rref#1{(\ref{#1})}
\newcounter{mycount} % counter for special list environment
\newenvironment{customlist}
  {\begin{list}
    {\arabic{mycount}}
    {\usecounter{mycount}
     \setlength{\labelwidth}{3em}
     \setlength{\labelsep}{4pt}
     \setlength{\itemsep}{2pt}
     \setlength{\leftmargin}{0.75cm}
     \setlength{\rightmargin}{0cm}
     \setlength{\itemindent}{0em}
    }
  }
{\end{list}}

\def\shuff#1#2{\mathbin{
      \hbox{\vbox{\hbox{\vrule \hskip#2 \vrule height#1 width 0pt}\hrule}\vbox{\hbox{\vrule \hskip#2 \vrule height#1 width 0pt\vrule }\hrule}}}}
\def\shuffl{{\mathchoice{\shuff{5pt}{3.5pt}}{\shuff{5pt}{3.5pt}}{\shuff{3pt}{2.6pt}}{\shuff{3pt}{2.6pt}}}}
\def\shuffle{{\, \shuffl \,}}
\def\allpolyprop{\mbox{$\re_{p}\langle X \rangle$}}
\def\allseriesnp{\mbox{$\re_{np}\langle\langle X \rangle\rangle$}}
\def\allseriesprop{\mbox{$\re_{p}\langle\langle X \rangle\rangle$}}

\begin{document}

\begin{frontmatter}

\title{Decompositions of Nonlinear Input-Output Systems to Zero the Output}

\author[odu]{W.~Steven Gray\corref{cor}}
\ead{sgray@odu.edu}

\author[ntnu]{Kurusch Ebrahimi-Fard}
\ead{kurusch.ebrahimi-fard@ntnu.no}

\author[ntnu]{Alexander Schmeding}
\ead{alexander.schmeding@ntnu.no}

\cortext[cor]{Corresponding author}

\address[odu]{Department of Electrical and Computer Engineering, Old Dominion University, Norfolk, Virginia 23529, USA}
\address[ntnu]{Department of Mathematical Sciences,
		Norwegian University of Science and Technology (NTNU),
		7491 Trondheim, Norway}

\begin{abstract}
Consider an input-output system where the output is the tracking error given some desired reference signal. It is natural to consider under what conditions the problem has an exact solution, that is, the tracking error is exactly the zero function. If the system has a well defined relative degree and the zero function is in the range of the input-output map, then it is well known that the system is locally left invertible, and thus, the problem has a unique exact solution. 
A system will fail to have relative degree when more than one exact solution exists.
The general goal of this paper is to describe a decomposition of an input-output system having a Chen-Fliess series representation into a parallel product of subsystems 
in order to identify possible solutions to the problem of zeroing the output.
For computational purposes, the focus is on systems whose generating series are polynomials.
It is shown that the shuffle algebra on the set of generating polynomials is a unique factorization domain so that
any polynomial can be uniquely factored modulo a permutation into its irreducible elements for the purpose
of identifying the subsystems in a parallel product decomposition.
This is achieved using the fact that this shuffle algebra is isomorphic to
the symmetric algebra over the vector space spanned by Lyndon words.
A specific algorithm for factoring generating polynomials into its irreducible factors
is presented based on the Chen-Fox-Lyndon factorization of words.
\end{abstract}

\begin{keyword}
nonlinear control systems, Chen-Fliess series, shuffle algebra
\end{keyword}

\end{frontmatter}

\section{Introduction}

Consider a smooth control-affine state space realization
\begin{subequations} \label{eq:SISO-realization}
\begin{align}
\dot{z}&=g_0(z)+g_1(z)u,\;\;z(0)=z_0 \label{eq:state}\\
y&=h(z),
\end{align}
\end{subequations}
where $g_0$, $g_1$, and $h$ are defined on $W\subseteq \re^n$.
If the realization has a well defined
relative degree at $z_0\in W$, then it is a classical result that the corresponding single-input, single-output map
$F:u\mapsto y$ is left invertible on a neighborhood of $z_0$ \cite{Isidori_95,Nijmeijer-vanderSchaft_90}. If the zero output is known to be in the range of $F$ for
some class of inputs $\cal U$, then there exists a unique input $u^\ast\in \cal U$ satisfying $F[u^\ast]=0$ which
can be generated in real-time using feedback \cite{Isidori_95,Nijmeijer-vanderSchaft_90} or computed analytically using formal
power series methods \cite{Gray-etal_Automatic_14}. This construction leads to the notion of {\em zero dynamics}
\cite{Gray-etal_CISS21,Isidori_95,Isidori_13,Nijmeijer-vanderSchaft_90}
and has well known applications in output tracking
and path planning problems when the output is taken to be an error signal \cite{Gray-etal_Automatic_14,Isidori_13,Nijmeijer-vanderSchaft_90}.
The problem of zeroing the output can also be applied in optimal control problems
in order to determine critical points of functional derivatives \cite{Duffaut-Espinosa-etal_CDC23}.

\begin{figure}[tb]
\begin{center}
\includegraphics[scale=0.4]{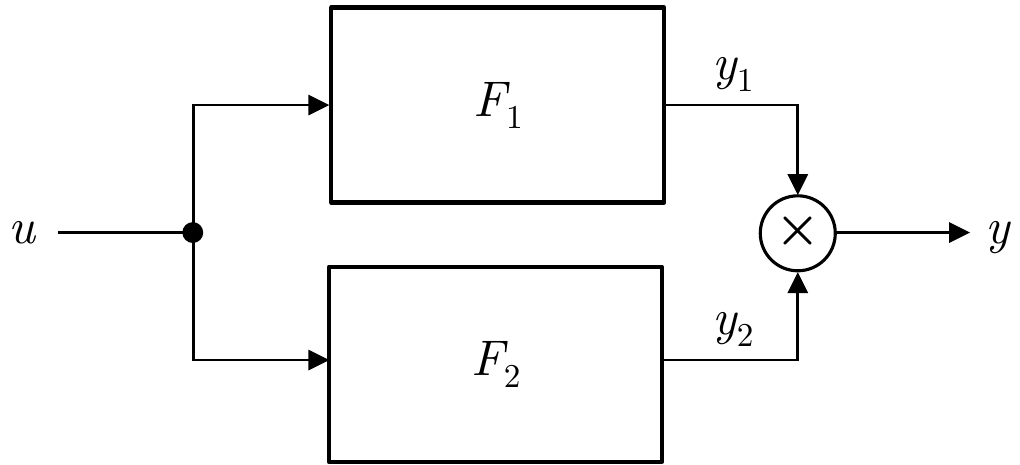}
\caption{Parallel product connection of systems $F_1$ and $F_2$}
\label{fig:parallel-product-F1-F2}
\end{center}
\end{figure}

System \rref{eq:SISO-realization} can fail to have relative degree under a number of different circumstances.
For example, it can turn out that the Lie derivative $L_{g_1}h$ is exactly zero at $z_0$.
As explained in \cite[Remark 3]{Tanwani-Liberzon_10}, it is still possible for the system to be
left invertible in a certain sense if the state trajectory
immediately leaves the singularity after $t=0$, and two inputs are said to be equivalent when they differ only on a set of measure zero.
Another possibility is the case where the input-output map is simply not left-invertible.
Take as a simple example the system
\begin{subequations} \label{eq:intro-example-two-factors}
\begin{align}
\dot{z}_1&=1-u,\;\;
\dot{z}_2=z_3-u,\;\;
\dot{z}_3=1,\;\;
z(0)=0 \\
y&=z_1z_2.
\end{align}
\end{subequations}
It is easily verified that this realization does not have relative degree at the origin.
The input-output map is not
injective since
there are two inputs which give the zero output: $u_1^\ast(t)=1$, $t\geq 0$ and $u_2^\ast(t)=t$, $t\geq 0$.
This is a result of the fact that the system can be decomposed into a parallel product of two subsystems
$F_1:u\mapsto y_1=z_1$ and $F_2:u\mapsto y_2=z_2$
as shown in Figure~\ref{fig:parallel-product-F1-F2}, where each
subsystems has the zero function in its range.
The general goal of this
paper is to
describe
how to perform this decomposition in order to identify possible solutions to the problem of zeroing the output.
This problem is only nontrivial in the nonlinear setting since linear time-invariant systems always have relative degree.

The approach taken will be to work purely in the input-output setting using Chen-Fliess series representations.
One advantage to this point of view is that the nonuniqueness of coordinate systems can be avoided.
That is, the generating series for the input-output map of a state space realization is invariant under
coordinate transformation.
In addition, this framework is more general as every analytic state space realization \rref{eq:SISO-realization} has an input-output map
with a Chen-Fliess series representation but not conversely.
In order to avoid convergence issues associated with such series, the analysis will be done using
{\em formal} Fliess operators \cite{Gray-Wang_08}, that is, maps that take an infinite jet representing a
formal input function to an infinite jet representing a formal output function.
In this context, the problem of zeroing the output boils down to a purely algebraic problem.

The concept of a {\em nullable} generating series is presented first. This is a formal power series representing
a formal Fliess operator having the property that the zero output (jet) is in the range of the operator.
A generating series is called {\em strongly nullable} if there is a nonzero input that maps to the zero output
and {\em primely nullable} if this input is the only input with this property.
A special class of primely nullable series are those having relative degree and one additional property. These will be called {\em linearly nullable}.
While there is no known general test for nullability, linearly nullable series can be completely characterized,
and their nulling inputs can be computed directly using the methods in \cite{Gray-etal_Automatic_14}.
A key fact is that the parallel product of two Chen-Fliess series has a Chen-Fliess series representation whose generating series is the shuffle product of the generating series of the subsystems \cite{Fliess_81}. It will be shown that the shuffle product
of two linearly nullable series is always strongly nullable
but never linearly nullable. On the other hand,
a given series may not be linearly nullable but if any of its shuffle
factors is linearly nullable, then the problem of zeroing the output can be directly solved since nulling one factor will
zero the system output.
The main issue then becomes how to factor a generating series
into its irreducible elements in the shuffle algebra.

The algorithmic focus in this paper will be on the polynomial case.
It is first established that the shuffle algebra as a
commutative polynomial ring over a finite set of noncommuting indeterminates is a unique factorization domain. This is achieved by assembling existing results
from algebra \cite{Cohn_73} and algebraic combinatorics \cite{Lothaire_83}. Of particular importance is the fact that this
shuffle algebra can be viewed as the symmetric algebra over the $\re$-vector space spanned by Lyndon words
\cite{Lothaire_83,Radford_79}.
Next, an algorithm is given to factor a polynomial into its irreducible shuffle components. This is done by first mapping the polynomial
to the symmetric algebra using the Chen-Fox-Lyndon factorization of words
\cite{Chen-etal_58,Hazewinkel_01,Lothaire_83,Radford_79,Reutenauer_93}. The resulting polynomial is then factored using one of the
many known algorithms for factoring multivariate commutative polynomials \cite{von_zur_Gathen_85}. Then each factor is
mapped back to the shuffle algebra.

The paper is organized as follows. In the next section, a brief summary is given of the mathematical
tools used to establish the main results of the paper. In Section~\ref{sec:nullable-gs}, the concept of nullable generating series
is presented. The subsequent section addresses the problem of factoring generating series in the shuffle algebra.
The final section provides the main conclusions of the paper.
It should be stated that a shorter preliminary version of this paper was presented as a conference paper \cite{Gray-etal_CISS23}.
The present version includes three additional examples
(Examples~\ref{ex:nulling-maximal-series}, \ref{ex:nulling-optimal-control}, and \ref{ex:mathscrL}),
a proof of identity \rref{eq:left-shift-composition-product} (see appendix),
a new Corollary~\ref{co:affine-subspace},
a revised and expanded version of the
proof of Theorem~\ref{th:shuffle-algebra-not-PID}, an expanded description of the factorization algorithm in Section~\ref{sec:factorization},
two new figures,
and seven additional references. In particular, Example~\ref{ex:nulling-optimal-control} provides a new application of the main results in the context of optimal control, while the revised proof of Theorem~\ref{th:shuffle-algebra-not-PID} is simpler and more direct.

\section{Preliminaries}

An {\em alphabet} $X=\{ x_0,x_1,$ $\ldots,x_m\}$ is any nonempty and finite set
of symbols referred to as {\em
letters}. A {\em word} $\eta=x_{i_1}\cdots x_{i_k}$ is a finite sequence of letters from $X$.
The number of letters in a word $\eta$, written as $\abs{\eta}$, is called its {\em length}.
The empty word, $\emptyset$, is taken to have length zero.
The collection of all words having length $k$ is denoted by
$X^k$. Define $X^\ast=\bigcup_{k\geq 0} X^k$,
which is a monoid under the concatenation product.
Any mapping $c:X^\ast\rightarrow
\re^\ell$ is called a {\em formal power series}.
Often $c$ is
written as the formal sum $c=\sum_{\eta\in X^\ast}(c,\eta)\eta$,
where the {\em coefficient} $(c,\eta)\in\re^\ell$ is the image of
$\eta\in X^\ast$ under $c$.
The {\em support} of $c$, $\supp(c)$, is the set of all words having nonzero coefficients.
A series $c$ is called {\em proper} if $\emptyset\notin\supp(c)$.
The {\em order} of $c$, $\ord(c)$, is the length of the shortest word in its support.
By definition the order of the zero series is $+\infty$.
The set of all noncommutative formal power series over the alphabet $X$ is
denoted by $\allseriesell$. The subset of series with finite support, i.e., polynomials,
is represented by $\allpolyell$.
Each set is an associative $\re$-algebra under the concatenation product and an associative and commutative $\re$-algebra under
the {\em shuffle product}, that is, the bilinear product uniquely specified by the shuffle product of two words
\begdi
	(x_i\eta)\shuffle(x_j\xi)=x_i(\eta\shuffle(x_j\xi))+x_j((x_i\eta)\shuffle \xi),
\enddi
where $x_i,x_j\in X$, $\eta,\xi\in X^\ast$ and with $\eta\shuffle\emptyset=\emptyset\shuffle\eta=\eta$ \cite{Fliess_81}.
For any letter $x_i\in X$, let $x_i^{-1}$ denote the $\re$-linear {\em left-shift operator} defined by $x_i^{-1}(\eta)=\eta^\prime$
when $\eta=x_i\eta^\prime$ and zero otherwise.
Higher order shifts are defined inductively via
$(x_i\xi)^{-1}(\cdot)=\xi^{-1}x_i^{-1}(\cdot)$, where $\xi\in X^\ast$. It acts as a derivation on the shuffle product.

\subsection{Chen-Fliess series}

Given any $c\in\allseriesell$ one can associate a causal
$m$-input, $\ell$-output operator, $F_c$, in the following manner.
Let $\mathfrak{p}\ge 1$ and $t_0 < t_1$ be given. For a Lebesgue measurable
function $u: [t_0,t_1] \rightarrow\re^m$, define
$\norm{u}_{\mathfrak{p}}=\max\{\norm{u_i}_{\mathfrak{p}}: \ 1\le
i\le m\}$, where $\norm{u_i}_{\mathfrak{p}}$ is the usual
$L_{\mathfrak{p}}$-norm for a measurable real-valued function,
$u_i$, defined on $[t_0,t_1]$.  Let $L^m_{\mathfrak{p}}[t_0,t_1]$
denote the set of all measurable functions defined on $[t_0,t_1]$
having a finite $\norm{\cdot}_{\mathfrak{p}}$ norm and
$B_{\mathfrak{p}}^m(R)[t_0,t_1]:=\{u\in
L_{\mathfrak{p}}^m[t_0,t_1]:\norm{u}_{\mathfrak{p}}\leq R\}$.
Assume $C[t_0,t_1]$
is the subset of continuous functions in $L_{1}^m[t_0,t_1]$. Define
inductively for each word $\eta=x_i\bar{\eta}\in X^{\ast}$ the map $E_\eta:
L_1^m[t_0, t_1]\rightarrow C[t_0, t_1]$ by setting
$E_\emptyset[u]=1$ and letting
\[E_{x_i\bar{\eta}}[u](t,t_0) =
\int_{t_0}^tu_{i}(\tau)E_{\bar{\eta}}[u](\tau,t_0)\,d\tau, \] where
$x_i\in X$, $\bar{\eta}\in X^{\ast}$, and $u_0=1$. The
{\em Chen-Fliess series} corresponding to $c\in\allseriesell$
is defined in \cite{Fliess_81} as
\begdi
y(t)=F_c[u](t) =
\sum_{\eta\in X^{\ast}} (c,\eta) \,E_\eta[u](t,t_0).
\enddi
If there exist real numbers $K_c,M_c>0$ such that
\begeq
\abs{(c,\eta)}\le K_c M_c^{|\eta|}|\eta|!,\;\; \forall\eta\in X^{\ast},
\label{eq:local-convergence-growth-bound}
\endeq
then $F_c$ constitutes a well defined mapping from
$B_{\mathfrak p}^m(R)[t_0,$ $t_0+T]$ into $B_{\mathfrak
q}^{\ell}(S)[t_0, \, t_0+T]$ for sufficiently small $R,T >0$ and some $S>0$,
where the numbers $\mathfrak{p},\mathfrak{q}\in[1,\infty]$ are
conjugate exponents, i.e., $1/\mathfrak{p}+1/\mathfrak{q}=1$  \cite{Gray-Wang_02}.
(Here, $\abs{z}:=\max_i \abs{z_i}$ when $z\in\re^\ell$.)
Any series $c$ satisfying \rref{eq:local-convergence-growth-bound} is called
{\em locally convergent}, and $F_c$ is called a {\em Fliess operator}.
The subset of all locally convergent series is denoted by $\allseriesellLC$.

A Fliess operator $F_c$ defined on $B^m_{\mathfrak p}(R)[t_0,t_0+T]$ with $\ell=1$
is said to be {\em realizable} when there exists
a state space realization \rref{eq:SISO-realization} with
each $g_i$ being an analytic vector field expressed in local
coordinates on some neighborhood ${W}$ of $z_0\in\re^n$,
and the real-valued output function $h$ is an analytic function on ${W}$ such that
(\ref{eq:state}) has a well defined solution $z(t)$, $t\in[t_0,t_0+T]$ for
any given input $u\in B^m_{\mathfrak p}(R)[t_0,t_0+T]$, and $y(t)=F_{c}[u](t)=h(z(t))$, $t\in[t_0,t_0+T]$.
Denoting the {\em Lie derivative} of $h$ with respect to $g_i$ by $L_{g_i}h$,
it can be shown that for any word $\eta=x_{i_k}\cdots x_{i_1}\in X^\ast$
\begeq \label{eq:realization2series}
(c,\eta)=L_{g_{\eta}}h(z_0):=L_{g_{i_1}}\cdots L_{g_{i_k}}h(z_0)
\endeq
\cite{Fliess_81,Isidori_95,Nijmeijer-vanderSchaft_90}.

\subsection{System interconnections}

Given Fliess operators $F_c$ and $F_d$, where $c,d\in\allseriesellLC$,
the parallel and product connections satisfy $F_c+F_d=F_{c+d}$ and $F_cF_d=F_{c\shuffle d}$,
respectively \cite{Fliess_81}. It is also known that the composition of two Fliess operators
$F_c$ and $F_d$ with $c\in\allseriesellLC$ and $d\in\allseriesmLC$ always yields another Fliess operator with generating series $c\circ d$, where the {\em composition product}
is given by
\begeq \label{eq:c-circ-d}
c\circ d=\sum_{\eta\in X^\ast} (c,\eta)\,\psi_d(\eta)(\mathbf{1})
\endeq
\cite{Ferfera_80}. Here
$\psi_d$ is the continuous (in the ultrametric sense) algebra homomorphism
from $\allseries$ to the vector space endomorphisms on $\allseries$, $\Endallseries$, uniquely specified by
$\psi_d(x_i\eta)=\psi_d(x_i)\circ \psi_d(\eta)$ with
$
\psi_d(x_i)(e)=x_0(d_i\shuffle e),
$
$i=0,1,\ldots,m$
for any $e\in\allseries$,
and where $d_i$ is the $i$-th component series of $d$
($d_0:=\mbf{1}:=1\emptyset$). By definition,
$\psi_d(\emptyset)$ is the identity map on $\allseries$.
The left-shift operators distribute over the composition product as follows:
\begeq \label{eq:left-shift-composition-product}
x_j^{-1}(c\circ d)=
\left\{\!\!\!
\begin{array}{ccl}
\displaystyle{x_0^{-1}(c)\circ d + \sum_{i=1}^m \!d_i\shuffle(x_i^{-1}(c)\circ d)}\!\!\!\!\!&:&\!\!\!\! \!j=0 \\
0\!\!\!\!\!&:&\!\!\!\! \!j\neq0.
\end{array}
\right.
\endeq
(See the appendix for a proof of this property.)
If $c,d\in\allseries$ with $m=\ell=1$ and $d$ non-proper, then one can define the quotient
$c/d=c\shuffle d^{\shuffle -1}$ so that $F_c/F_d=F_{c/d}$
with the shuffle inverse of $d$ defined as
\begdi
d^{\shuffle -1}=((d,\emptyset)(1-d^\prime))^{\shuffle -1}=(d,\emptyset)^{-1}(d^{\prime})^{\shuffle\ast},
\enddi
where $d^\prime=\mbf{1}-(d/(d,\emptyset))$ is proper and $(d^\prime)^{\shuffle\ast}:=\sum_{k\geq 0} (d^\prime)^{\shuffle k}$
\cite{Gray-etal_Automatic_14}.
The following lemma will be useful.

\begle \label{le:(c-div-d)-compose-e}
For any $c,d,e\in\allseries$ with $d$ non-proper, the following identity holds
\begdi
(c/d)\circ e=(c\circ e)/(d\circ e).
\enddi
\endle

\begpr
It can be shown directly from the definition of the composition product that if $d$ is non-proper then so is $d\circ e$.
In fact, $(d\circ e,\emptyset)=(d,\emptyset)\neq 0$.
Thus, both sides of the
equality in question are at least well defined formal power series. In light of the known identity
\begeq \label{eq:circ-action-on-shuffle}
(c\shuffle d)\circ e=(c\circ e)\shuffle(d\circ e)
\endeq
for any $c,d,e\in\allseries$ \cite{Foissy_15}, it is sufficient to show that
\begeq \label{eq:circ-action-on-shuffle-inverse}
d^{\shuffle -1}\circ e=(d\circ e)^{\shuffle -1}.
\endeq
It is clear via induction that for any $k\in\nat$,
\begdi
d^{\shuffle k}\circ e=(d\circ e)^{\shuffle k}.
\enddi
Therefore, since $d$ is non-proper, it follows that
\begin{align*}
d^{\shuffle -1}\circ e&=(d,\emptyset)^{-1}\lim_{n\rightarrow \infty}\sum_{k=0}^n(d^\prime)^{\shuffle k}\circ e \\
&=(d\circ e,\emptyset)^{-1}\lim_{n\rightarrow\infty}\sum_{k=0}^n(d^\prime\circ e)^{\shuffle k}\\
&=(d\circ e)^{\shuffle -1}.
\end{align*}
As $d^\prime$ and $d^\prime\circ e$ are both proper, all the limits above (in the
ultrametric sense) exist, and thus, the claim is verified.
\endpr

\subsection{Formal Fliess operators}

Suppose $X=\{x_0,x_1\}$, $X_0=\{x_0\}$, and define $\allseriesXO$ to be the set of all (commutative)
formal power series in the letter $x_0$.
Then every series $c_u\in\allseriesXO$ can be identified with an infinite jet $j^\infty_{t_0}(u)$ for
any fixed $t_0\in\re$. By Borel's Lemma, there is a real-valued function $u\in C^\infty(t_0)$ whose Taylor series corresponds
to $j^\infty_{t_0}(u)$. In the event that the coefficients of $c_u$ satisfy the growth bound \rref{eq:local-convergence-growth-bound},
then $u$ is real analytic. In which case, for any $c\in\allseriesLC$, $F_{c_y}[v]=y=F_c[u]=F_c[F_{c_u}[v]]=F_{c\circ c_u}[v]$, where $v$ is just a placeholder
in this chain of equalities. If
the Taylor series for $u$ does not converge, it is viewed as a formal function. Nevertheless, the
mapping $c\,\circ:\allseriesXO\rightarrow\allseriesXO:c_u\mapsto c_y=c\circ c_u$ is still well defined and takes the input infinite jet
to the output infinite jet. This is called a {\em formal Fliess operator} \cite{Gray-Wang_08}. The advantage of working
with these formal objects is that their algebraic properties can be characterized independently of their analytic nature.
This will be the approach taken below.

\subsection{Relative degree of a generating series}

Observe that
$c\in\allseries$ can always be decomposed into its natural and forced components, that
is, $c=c_N+c_F$, where $c_N:=\sum_{k\geq 0} (c,x_0^k)x_0^k$
and $c_F:=c-c_N$.

\begde\cite{Gray-etal_Automatic_14} \label{de:relative-degree-c}
Given $c\in\allseries$ with $X=\{x_0,x_1\}$,
let $r\geq 1$ be the largest integer such that $\supp(c_F)\subseteq x_0^{r-1}X^\ast$.
Then $c$ has {\em relative degree} $r$ if the linear word $x_0^{r-1}x_1\in \supp(c)$,
otherwise it is not well defined.
\endde

It is immediate that $c$ has relative degree $r$ if and only if
there exists some $e\in\allseries$ with $\supp(e)\subseteq X^\ast/\{X_0^\ast,x_1\}$ such that
\begeq \label{eq:cF-relative-degree-decomposition}
c=c_N+c_F=c_N+Kx_0^{r-1}x_1+x_0^{r-1}e
\endeq
and $K\neq 0$.
This notion of relative degree coincides
with the usual definition given in a state space
setting \cite{Gray-Ebrahimi-Fard_SIAM17}.

\section{Nullable Generating Series}
\label{sec:nullable-gs}

It is assumed for the remainder of the paper that all systems are single-input, single-output, i.e., $m=\ell=1$
so that $X=\{x_0,x_1\}$ and all series coefficients are real-valued.
Consider the following classes of generating series.

\begde
A series $c\in\allseries$ is said to be {\em nullable} if the zero series is in the range of the mapping
$c\,\circ:\allseriesXO\rightarrow\allseriesXO,c_u\mapsto c\circ c_u$.
That is, there exists a {\em nulling series} $c_{u^\ast}\in\allseriesXO$ such that $c\circ c_{u^\ast}=0$.
The series is  {\em strongly nullable} if it has a nonzero nulling series.
A strongly nullable series is {\em primely nullable} if its nulling series is unique.
\endde

Observe that from \rref{eq:c-circ-d} it follows that $(c\circ c_u,\emptyset)=(c,\emptyset)$ for all $c_u\in\allseriesXO$. Thus,
if $c$ is nullable, then necessarily $c$ must be proper.
Also, every series $c=c_F$ satisfies $c\circ 0=0$. Thus, it is nullable.
If $c=c_N+c_F$ with $c_N \neq 0$, then $c\circ 0=c_N$. Therefore, if
$c$ is nullable, it must be strongly nullable.

\begex \label{ex:no-r-but-nullable}
{\rm
Observe that $c=x_0^2-x_1x_0$ is primely nullable since $c\circ \mbf{1}=x_0^2-x_0^2=0$, and
$c_{u^\ast}=\mbf{1}$ is the only series with this property.
}
\endex

\begex \label{ex:has-r-but-nullable}
{\rm
The polynomial $c=x_0+x_0x_1$ is not nullable since $c\circ c_u=x_0+x_0^2c_u\neq 0$ for all $c_u\in\allseriesXO$.~\hfill\openbull
}
\end{myexample}

A sufficient (but not necessary) condition for a series to be primely nullable is given in the following theorem.
It has its roots in \cite{Gray-etal_Automatic_14}. A concise proof is given here to make
its computational features explicit and the subsequent corollary more apparent.

\begth \label{th:zero-invertible-implies-nullable}
If $c\in\allseries$ has relative degree $r$, and $\supp(c_N)\subseteq x_0^rX_0^\ast$ is nonempty, then $c$ is primely nullable.
\endth

\begpr
Since $c_N\neq 0$ by assumption, any nulling series must be nonzero. The claim is that $c$ has a unique nonzero nulling series.
Applying \rref{eq:left-shift-composition-product} to
$c_y=c\circ c_u$
with $m=1$ (let $d_1=d$)
under the assumption that $c$ has relative degree $r$ gives
\begin{align*}
c_y&=c\circ c_u \\
x_0^{-1}(c_y)&=x_0^{-1}(c)\circ c_u \\
&\hspace*{0.08in}\vdots \\
x_0^{-r+1}(c_y)&=x_0^{-r+1}(c)\circ c_u \\
x_0^{-r}(c_y)&=x_0^{-r}(c)\circ c_u+c_u\shuffle((x_0^{r-1}x_1)^{-1}(c)\circ c_u).
\end{align*}
Since $(x_0^{r-1}x_1)^{-1}(c)$ is non-proper (specifically, $((x_0^{r-1}x_1)^{-1}(c),\emptyset)=K\neq 0$
in \rref{eq:cF-relative-degree-decomposition}) it can be shown that $(x_0^{r-1}x_1)^{-1}(c)\circ c_u$ is also
non-proper and thus has a shuffle inverse. Setting  $x_0^{-r}(c_y)=0$ and dividing by $(x_0^{r-1}x_1)^{-1}(c)\circ c_u$
gives
\begdi
0=(x_0^{-r}(c)\circ c_u)/((x_0^{r-1}x_1)^{-1}(c)\circ c_u)+c_u.
\enddi
Next, applying Lemma~\ref{le:(c-div-d)-compose-e} yields
\begdi
0=(x_0^{-r}(c)/(x_0^{r-1}x_1)^{-1}(c))\circ c_u+c_u.
\enddi
Define a generalized series $\delta$ with the defining property that $F_{\delta}[u]=u$ for all admissible inputs $u$. Then it
must have the unital property $\delta\circ c=c\circ\delta=c$ on the semigroup
$(\allseries,\circ)$. The previous equation
can be written as
\begdi
0=\underbrace{(\delta+(x_0^{-r}(c)/(x_0^{r-1}x_1)^{-1}(c)))}_{:=d_\delta}\circ c_u.
\enddi
It is known that the set of series $\delta+\allseries$ forms a group under the induced composition product \cite{Gray-etal_SCL14}.
Therefore, one can solve for $c_u$ directly via left inversion to give
$
c_u=d_\delta^{\circ -1}\circ 0.
$
In which case, there exists a unique $c_u$ that will zero out all the coefficients of $c_y$ with the exception of the first $r$ coefficients.
These initial coefficients are completely determined by $c$ since
\begin{align*}
(c_y,x_0^k)&=(x_0^{-k}(c_y),\emptyset)=(x_0^{-k}(c)\circ c_u,\emptyset)\\
&=(x_0^{-k}(c),\emptyset)=(c,x_0^k),\;\;k=0,1,\ldots,r-1.
\end{align*}
By assumption $\supp(c_N)\subseteq x_0^rX_0^\ast$. Hence, all the coefficients above must be zero so that $c_y=0$ as desired.
\endpr

It is worth noting that $(\allseries,\circ,\delta)$ described above as well as $(\allseries,\shuffle,\mbf{1})$ both
include the monoids of characters over their respective graded connected bialgebras of coordinate functions. Identity \rref{eq:circ-action-on-shuffle}, which is
central in this work, can then be viewed in terms of the concept of two bialgebras in cointeraction \cite{Foissy_22}. In this respect, equation \rref{eq:circ-action-on-shuffle-inverse} is equivalent to stating that the right action of the
character monoid $(\allseries,\circ,\delta)$ on the group of unital non-proper
series $(\mbf{1} + \allseriesnp ,\shuffle,\mbf{1}) \subset (\allseries,\shuffle,\mbf{1})$ is compatible with
the antipode of its Hopf algebra of coordinate functions.

Series satisfying the condition in Theorem~\ref{th:zero-invertible-implies-nullable} will be referred to as {\em linearly nullable}
since the linear word $x_0^{r-1}x_1$ in its support plays a key role in computing the nulling series.
In light of \rref{eq:cF-relative-degree-decomposition}, every such series has the form
\begdi
c=x_0^re_0+Kx_0^{r-1}x_1+x_0^{r-1}e_1,
\enddi
where $r\in\nat$, $K\neq 0$, $e_0\in\allseriesXO/\{0\}$, and $\supp(e_1)\subseteq X^\ast/\{X_0^\ast,x_1\}$.
From the general identity $(x_0^kc)\circ d=x_0^k(c\circ d)$ and the
fact that the composition product is left linear, it follows that
\begin{align*}
c\circ c_{u^\ast}&=(x_0^re_0+Kx_0^{r-1}x_1+x_0^{r-1}e_1)\circ c_{u^\ast} \\
&=x_0^{r-1}((x_0e_0+Kx_1+e_1)\circ c_{u^\ast})\\
&=0.
\end{align*}
That is, $c_{u^\ast}$ is a solution to
\begdi
(x_0e_0+Kx_1+e_1)\circ c_{u^\ast}=x_0(e_0+Kc_{u^\ast})+e_1\circ c_{u^\ast}=0.
\enddi

Central to the proof of Theorem~\ref{th:zero-invertible-implies-nullable}
is the observation that mapping $c\circ$
under the condition that $c$ has relative degree is
injective since it is left invertible. The following corollary, which also follows directly
from this proof, states
that $c\circ$ is never surjective on $\allseriesXO$ under this condition.

\begco \label{co:affine-subspace}
Suppose $c$ has relative degree $r$. Define $c_N^{r-1}=(c,\emptyset)+(c,x_0)x_0+\cdots +(c,x_0^{r-1})x_0^{r-1}$. Then the range of the
mapping $c_u\mapsto c\circ c_u$ is the affine subspace of the $\re$-vector space $\allseriesXO$
\begdi
R_c:=\{c_y=c_N^{r-1}+x_0^r e:e\in\allseriesXO\}.
\enddi
Therefore, $c$ is nullable in this case only if $c^{r-1}_N=0$.
\endco

\begex \label{ex:nullability-x0+x1}
{\rm
The polynomial $c=x_0+x_1$ has relative degree 1 and $c_N=x_0\in x_0X_0^\ast$. Therefore, it is linearly nullable.
Specifically, $c_{u^\ast}=-\mbf{1}$ is the only series that yields $c\circ c_{u^\ast}=0$.
}
\endex

\begex
{\rm
The polynomial $c=x_0^2-x_1x_0$ in Example~\ref{ex:no-r-but-nullable} does not have relative degree
since it does not have a linear word in its support (i.e., $K=0$ in  \rref{eq:cF-relative-degree-decomposition}).
So it is primely nullable but not linearly nullable.
}
\endex

\begex
{\rm
The polynomial $c=x_0+x_0x_1$ in Example~\ref{ex:has-r-but-nullable} has relative degree 2 and was shown not to be nullable.
Observe $c_N=x_0\not\in x_0^2X_0^\ast$, which is
consistent with Theorem~\ref{th:zero-invertible-implies-nullable}.
}
\endex

\begex \label{ex:nulling-maximal-series}
{\rm
Consider the series $c=\sum_{\eta\in X^+}|\eta|!\,\eta$, where $X^+:=X^\ast/\{\emptyset\}$.
The series has relative degree 1 and is linearly nullable. In this instance, the corresponding
Chen-Fliess series has the closed-form expression
\begdi
F_c[u]=\frac{F_{x_0+x_1}[u]}{1-F_{x_0+x_1}[u]}.
\enddi
Therefore, the unique nulling series for $c$ is $c_{u^\ast}=-\mbf{1}$.
}
\endex

Let $c\in\allseries$ be nullable. Define the (two-sided) principal ideal
\begdi
I_c=(c):=\{c\shuffle d: d\in\allseries\}
\enddi
in the shuffle algebra on $\allseries$.

\begle \label{le:shuffle-ideal}
Every series in $I_c$ is nullable. If $c$ is strongly nullable, then every series in $I_c$ is strongly
nullable.
\endle

\begpr
Applying \rref{eq:circ-action-on-shuffle} it follows that $(c\shuffle d)\circ c_{u^\ast}=(c\circ c_{u^\ast})\shuffle (d\circ c_{u^\ast})=0$ if
$c_{u^\ast}$ is selected so that $c\circ c_{u^\ast}=0$, which is always possible since $c$ is nullable by assumption.
The second claim is now obvious.
\endpr

The first theorem below is obvious given the definition of primely nullable. The second theorem
is less trivial but not unexpected. It confirms that the set of linearly nullable series is not
closed under the shuffle product.

\begth \label{th:pnull-shuffle-pnull}
If $c,d\in\allseries$ are primely nullable with $c_{u^\ast}\neq d_{u^\ast}$,
then $c\shuffle d$ is strongly nullable but not primely nullable.
\endth

\begth \label{th:lnull-shuffle-lnull}
If $c,d\in\allseries$ are linearly nullable,
then $c\shuffle d$ is strongly nullable but not linearly nullable.
\endth

\begpr
The strong nullability property follows directly from the lemma above.
Regarding the second assertion, if
$c\shuffle d$ is linearly nullable, then necessarily $c\shuffle d$ must have
relative degree, say $s$, and $(c\shuffle d)_N\in x_0^sX_0^\ast$. Observe that
\begin{align*}
c\shuffle d&=(x_0^{r_c}e_0+K_cx_0^{r_c-1}x_1+x_0^{r_c-1}e_1)\shuffle \\
&\hspace*{0.2in}(x_0^{r_d}f_0+K_dx_0^{r_d-1}x_1+x_0^{r_d-1}f_1)
\end{align*}
has the property that $(c\shuffle d)_N\in x_0^{r_c+r_d}X_0^\ast$. But the assertion is that
$c\shuffle d$ cannot have relative degree $r_c+r_d$. This would require that
the shortest linear word in $\supp(c\shuffle d)_F$ be $x_0^{r_c+r_d-1}x_1$
and all other words in $\supp((c\shuffle d)_F)$
must have the prefix $x_0^{r_c+r_d-1}$. This linear word will only be present if
\begeq \label{eq:linear-word-coefficient}
K_c(f_0,\emptyset)+K_d(e_0,\emptyset)\neq 0.
\endeq
This means that at least one of the constant terms $(e_0,\emptyset)$ or $(f_0,\emptyset)$ must be nonzero.
In addition, note that every word in the support of
\begin{align*}
\lefteqn{(e_0,\emptyset)x_0^{r_c}\shuffle K_dx_0^{r_d-1}x_1+(f_0,\emptyset)x_0^{r_d}\shuffle K_cx_0^{r_c-1}x_1} \\
&=K_d(e_0,\emptyset)(x_0^{r_c}\shuffle x_0^{r_d-1}x_1)+K_c(f_0,\emptyset)(x_0^{r_d}\shuffle x_0^{r_c-1}x_1)
\end{align*}
has length $r_c+r_d$, and these words must have the required prefix $x_0^{r_c+r_d-1}$ since
no other words in the larger shuffle product are short enough to cancel these words. But the
only way to remove an illegal word would violate \rref{eq:linear-word-coefficient}.
For example, if $r_c=r_d=1$, then
\begin{align*}
\lefteqn{(e_0,\emptyset)x_0\shuffle K_dx_1+(f_0,\emptyset)x_0\shuffle K_cx_1} \\
&=K_d(e_0,\emptyset)(x_0x_1+x_1x_0)+K_c(f_0,\emptyset)(x_0x_1+x_1x_0).
\end{align*}
The illegal word $x_1x_0$ cannot be canceled without removing the required linear word $x_0x_1$.
Thus, $c\shuffle d$ cannot be linearly nullable.
\endpr

\begex
{\rm
Suppose $c=x_0-x_1$ and $d=x_0^2-x_1$. Both series are linearly nullable with relative degree 1.
The nulling series for $c$ is $c_{u^\ast}=\mbf{1}$, and
the nulling series for $d$ is $d_{u^\ast}=x_0$. Observe
\begdi
c\shuffle d=-x_0x_1 - x_1x_0 + 2 x_1^2 + 3 x_0^3 - x_0^2x_1 - x_0x_1x_0 - x_1x_0^2
\enddi
does not have relative degree. Therefore $c\shuffle d$ is strongly nullable, but not linearly nullable
and not primely nullable.
In fact, if the coefficients for the realization \rref{eq:intro-example-two-factors} are computed
from \rref{eq:realization2series}, one will find directly that the generating series is the polynomial given above.
This is the origin of the example given in the introduction.
}
\endex

\begex \label{ex:lnull-shuffle-non-null}
{\rm
Suppose $c=x_0+x_1$ and $d=\mbf{1}+x_1$. In this case, $c$ is linearly nullable with relative degree 1, and $d$ also has
relative degree 1 but is not nullable as it is not proper. Observe
\begdi
c\shuffle d=x_0+x_1+x_0x_1+x_1x_0+2x_1^2
\enddi
is also linearly nullable with relative degree 1. That is, Theorem~\ref{th:lnull-shuffle-lnull} does not preclude
the possibility that primely nullable series can have shuffle factors that are not nullable.
}
\endex

\begex \label{ex:lnull-shuffle-lnull-equal}
{\em
Suppose $c=d=x_0-x_1$ so that both series are linearly nullable with relative degree 1.
As expected,
\begdi
c\shuffle d=2x_0^2-2x_0x_1-2x_1x_0-2x_1^2
\enddi
is not linearly nullable as it does not have relative degree, but it is primely nullable since $c_{u^\ast}=d_{u^\ast}=\mbf{1}$ is the only nulling series for
$c\shuffle d$ as the shuffle product is an integral domain. That is, in general
$(c\shuffle d)\circ e_u=(c\circ e_u)\shuffle (d\circ e_u)=0$ if and only if at least one argument
in the second shuffle product is the zero series.
}
\endex

In summary, if $\allseriesprop$ is the set of all proper series in $\allseries$, then the following inclusions hold:
\begin{center}
\begin{minipage}{3in}
$\allseriesprop$ $\supset$ nullable series  $\supset$ strongly nullable series $\supset$ primely nullable series $\supset$ linearly nullable series.
\end{minipage}
\end{center}
In light of Theorems~\ref{th:pnull-shuffle-pnull} and \ref{th:lnull-shuffle-lnull},
only the set of nullable series and strongly nullable series are closed under the shuffle product.

The final example provides an application of nullable series.

\begex \label{ex:nulling-optimal-control}
\rm
In optimal control problems, it is often necessary to determine critical points of quadratic
objective functions. From the calculus of variations, this is accomplished by computing
the critical points of a variational derivative. For a system described only in terms of
a Chen-Fliess series, this would involve determining the critical points of a variational
derivative of a Chen-Fliess
series.
The Fr\'{e}chet derivative of $F_c$, for example, can be computed by
introducing a variational alphabet associated with $X=\{x_0,x_1\}$, say $\delta X=\{\delta x_0,\delta x_1\}$.
Define the mapping $\delta: X\rightarrow \delta X$ by
$\delta(x_0)=\delta x_0=0$ and $\delta(x_1)=\delta x_1$.
Extend the definition of $\delta$
to $X^\ast$ by letting it act as a derivation with respect to concatenation.
Further extend the definition to $\allseries$ by linearity.
In which case, the Fr\'{e}chet derivative of $F_c$ at $u$ is the linear functional
$DF_c[u][h]=F_{\delta(c)}[u,h]$ \cite{Duffaut-Espinosa-etal_CDC23}.
Consider the simple example where
$c=x_0x_1+x_1x_0+x_1^2$ so that
$\delta(c)=(x_0+x_1)\shuffle \delta x_1$.
Identifying $u$ with $x_1$ and $h$ with $\delta x_1$ from some admissible set of functions ${\cal U}$, it follows that
\begdi
DF_c[u][h]=F_{x_0+x_1}[u]E_{\delta x_1}[h],
\enddi
where $DF_c[u][\cdot]$ is clearly linear.
Critical points in this context are the inputs $u^\ast\in\cal U$ such that $DF_c[u^\ast][h]=0$ for all $h\in\cal U$.
Here it is evident since $x_0+x_1$ is linearly nullable that $u^\ast(t)=-1$, $t\geq 0$.
\endex

\section{Factorizations in the Shuffle Algebra}
\label{sec:factorization}

The shuffle product on $\allpoly$ forms a commutative ring.
Such structures appear in the following chain of class inclusions:

\begin{center}
\begin{minipage}{3in}
commutative rings $\supset$ integral domains $\supset$ integrally closed domains $\supset$
GCD domains $\supset$ \underline{unique factorization domains} $\supset$ principal
ideal domains $\supset$ Euclidean domains $\supset$ fields
\end{minipage}
\end{center}
\cite{Artin_11,Lombardi-Quitte_15}.
The integral domain property of the shuffle algebra was proved in \cite[Theorem 3.2]{Ree_58}.
The following theorem identifies the strongest structure available on this ring.

\begth \label{th:shuffle-algebra-not-PID}
The shuffle algebra on $\allpoly$ is a unique factorization domain but not a principal ideal domain.
\endth

\begpr
The claim that the shuffle algebra on $\allpoly$ is a unique factorization domain follows from existing results.
It is known from \cite{Radford_79} (see also \cite[Section 6]{Grinberg-Reiner_00} and \cite[Chapter 5]{Lothaire_83}) that the shuffle algebra on $\allpoly$
is isomorphic to the symmetric algebra on the
$\re$-vector space $V$ having basis $L=\{l_i\}_{i\geq 0}$, the set of Lyndon words. The symmetric algebra $S(V)$ is in turn canonically isomorphic
to the free polynomial algebra $\re[L]$.
Thus, there exists an $\re$-linear map ${\mathscr L}:\allpoly\rightarrow\re[L]$ such that
\begeq \label{eq:shuffle-isomorphism}
{\mathscr L}(\eta_1\shuffle\eta_2)={\mathscr L}(\eta_1){\mathscr L}(\eta_2),\;\;\forall \eta_i\in X^\ast
\endeq
with ${\mathscr L}(\mathbf{1})=\mathbf{1}$, and, in particular, ${\mathscr L}(l_i)=l_i$.
Put another way, the shuffle algebra on $\allpoly$ is freely generated by the set of all Lyndon words \cite[Theorem 6.1]{Reutenauer_93}.
It is shown in \cite[Section 4, Corollary 1]{Cohn_73} that any such polynomial ring is
a unique factorization domain (see also \cite{Berstel-Boasson_02}).

To be a principal ideal domain, it is necessary that every ideal in $\allpoly$ be generated by
a single element.
The classical argument that this is not the case in the present context goes as follows (e.g., see \cite[p.~153]{MacLane-Birkhoff_67}).
The assertion is that the set of all proper polynomials in $\allpoly$, $\allpolyprop$, is an ideal which is not principal.
It is clear that $\allpolyprop$ is an ideal. Now suppose $\allpolyprop$ has a single generator
$p$ in the shuffle algebra, i.e., $\allpolyprop=(p):=\{p\shuffle q: q\in\allpoly\}$.
Since $x_0,x_1\in\allpolyprop$, there must exist $q_0,q_1\in\allpoly$ such that
$x_0=p\shuffle q_0$ and $x_1=p\shuffle q_1$. In light of the degrees of $x_0$ and $x_1$, this
would require a generator of the form $p=\alpha_0x_0+\alpha_1 x_1$, $\alpha_i\in\re$. If $\alpha_0=0$,
then $p$ will generate $x_1$ but not $x_0$. Likewise, if $\alpha_1=0$ then $p$ will generate $x_0$ but not $x_1$.
Thus, the ideal $\allpolyprop$ has two
basis elements, that is, $\allpolyprop=(x_0,x_1):=\{x_0\shuffle q_0+x_1\shuffle q_1:q_0,q_1\in\allpoly\}$,
and thus is not principal.
\endpr

\begin{figure}[tb]
\begin{center}
\includegraphics[scale=0.4]{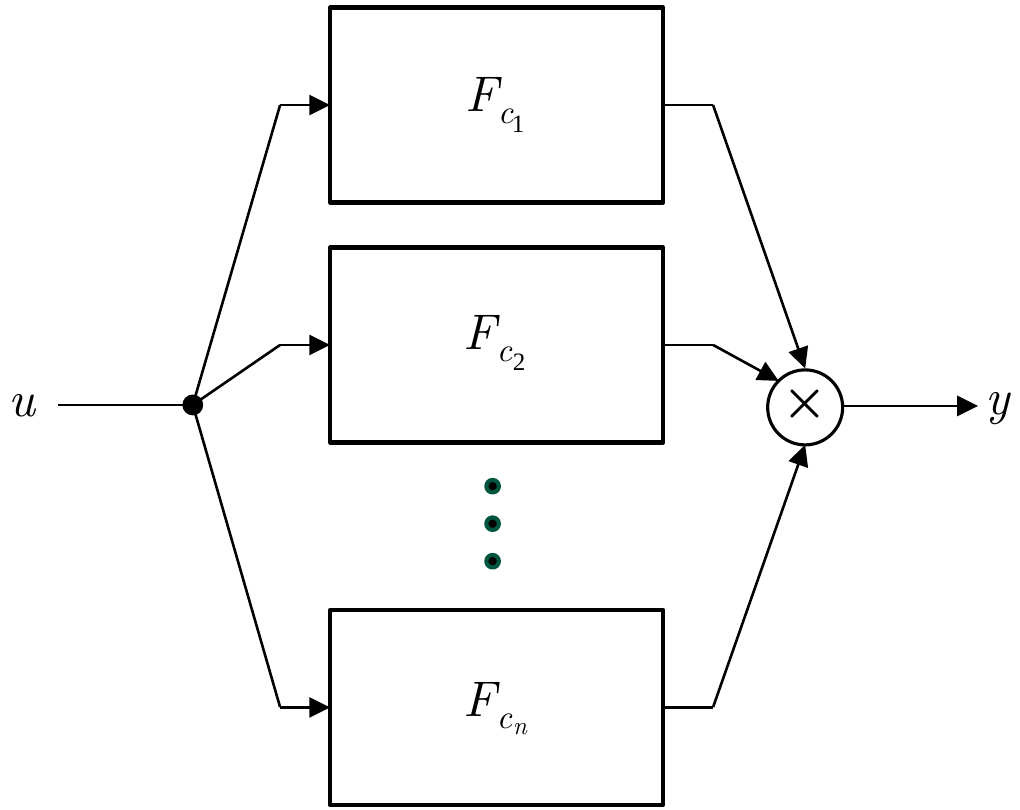}
\caption{Parallel product decomposition of $F_c$}
\label{fig:parallel-product-decomposition-Fc}
\end{center}
\end{figure}

The main theorem of this section is presented next.
It describes how to zero the output of a formal Fliess operator $F_c$ using its parallel product decomposition as shown in Figure~\ref{fig:parallel-product-decomposition-Fc}.

\begth
Let $c\in\allpoly$ with $c_N\neq 0$ and unique factorization $c=c_1\shuffle c_2\shuffle \cdots\shuffle c_n$ (modulo a permutation),
where each $c_i$ is irreducible as a polynomial in
the shuffle algebra.
Then $c_{u^\ast}\neq 0$ is a nulling series for $c$ if and only if it is a nulling series for at least one of the factors $c_i$.
\endth

\begpr
If $c_{u^\ast}\neq 0$ is a nulling series for $c_i$, then directly from Lemma~\ref{le:shuffle-ideal} it is a nulling series for $c$.
Conversely, if
$$c\circ c_{u^\ast}=(c_1\circ c_{u^\ast})\shuffle (c_2\circ c_{u^\ast})\shuffle \cdots\shuffle (c_n\circ c_{u^\ast})=0$$
for some
$c_{u^\ast}\neq 0$, then since the shuffle algebra is an
integral domain, at least one series $c_i\circ c_{u^\ast}$ must be the zero series, and the theorem is proved.
\endpr

It is important to point out what the theorem above is not saying, namely, that every nullable series can be factored into a
shuffle product of {\em primely} nullable series. While it is easy to demonstrate that a primely nullable series need not
be irreducible (Examples~\ref{ex:lnull-shuffle-non-null} and \ref{ex:lnull-shuffle-non-null-v2}),
it is unknown at present whether a nullable and irreducible series is always
primely nullable. This is a much deeper problem.

Next, an algorithm is given to shuffle
factorize any polynomial $c\in\allpoly$ into its irreducible components. This result
follows from combining the proof of the previous theorem and various known properties
of Lyndon words.
Suppose $X=\{x_0,x_1\}$ is ordered with $x_0<x_1$ so as to induce
a corresponding lexicographical ordering on $X^+$.
Recall that a word $\eta\in X^+$ is called a {\em Lyndon word}
if all factorizations $\eta=\xi\nu$ with $\xi,\nu\in X^+$ have the property that $\eta<\nu\xi$.
In this case, the first few Lyndon words are
$L=\{l_i\}_{i\geq 0}=\{x_0,x_1,x_0x_1, x_0^2x_1,x_0x_1^2, x_0^3x_1, x_0^2x_1^2,x_0x_1^3,\ldots\}$,
where here the ordering is by increasing word length and then lexicographically among words of the same
length.\footnote{This ordering is only for convenience in displaying results and does not play any mathematical role in this presentation.}
The Chen-Fox-Lyndon factorization of a word $\eta\in X^+$\
is a unique non-increasing product of Lyndon words so that
\begdi
\eta=l_{i_1}l_{i_2}\cdots l_{i_n},\;\;l_{i_1}\geq l_{i_2}\geq\cdots\geq l_{i_n}
\enddi
\cite{Chen-etal_58,Hazewinkel_01,Lothaire_83,Reutenauer_93}. A consequence of this factorization is that
\begeq \label{eq:Hazewinkel-decomposition}
l_{i_1}\shuffle l_{i_2}\shuffle\cdots \shuffle l_{i_n}=a\eta+R,
\endeq
where $a$ is a nonzero element in the set of rational numbers, $Q$, and all the words in the remainder $R$ are lexicographically smaller
than $\eta$ \cite[Theorem 5.6]{Hazewinkel_01}.
Applying \rref{eq:shuffle-isomorphism} to both sides of \rref{eq:Hazewinkel-decomposition} provides the recursive
formula
\begeq \label{eq:mathscrL}
\mathscr{L}(l_{i_1}l_{i_2}\cdots l_{i_n})=\frac{1}{a}(l_{i_1}l_{i_2}\cdots l_{i_n}-\mathscr{L}(R)).
\endeq
The recursion will terminate when all the remainders are zero.

\begex \label{ex:mathscrL}
{\rm
Let $\eta_1=x_0x_1x_0$, $\eta_2=x_0^2x_1x_0$, $\eta_3=x_0x_1x_0^2$. Their Chen-Fox-Lyndon factorizations are, respectively,
$\eta_1=(x_0x_1)x_0=l_2l_0$, $\eta_2=(x_0^2x_1)(x_0)=l_3l_0$, and  $\eta_3=(x_0x_1)(x_0)(x_0)=l_2l_0^2$.
Observe, for example, that
\begdi
l_2\shuffle l_0=x_0x_1x_0+2x_0^2x_1=\eta_1+R,
\enddi
where $R=2l_3$.
Therefore, from \rref{eq:mathscrL}
\begdi
\mathscr{L}(\eta_1)=l_0l_2-2l_3.
\enddi
Likewise,
\begin{align*}
\mathscr{L}(\eta_2)&=l_0l_3-3l_5 \\
\mathscr{L}(\eta_3)&=\frac{1}{2}l_0^2l_2-2l_0l_3+3l_5.
\end{align*}
To illustrate the isomorphism between the shuffle algebra on $\allpoly$ and the free algebra $R[L]$, observe that
\begdi
\eta_1\shuffle x_0=2\eta_2+2\eta_3,
\enddi
so that as expected from \rref{eq:shuffle-isomorphism}
\begin{align*}
\mathscr{L}(\eta_1\shuffle x_0)&=2\mathscr{L}(\eta_2)+2\mathscr{L}(\eta_3)\\
&=(l_0l_2-2l_3)l_0 \\
&=\mathscr{L}(\eta_1)\mathscr{L}(x_0).
\end{align*}
}
\endex

The proposed algorithm for shuffle factoring a proper polynomial $c$ is summarized below:
\begin{customlist}
\item[1.] Compute $c_L={\mathscr L}(c)$ using
\rref{eq:Hazewinkel-decomposition}-\rref{eq:mathscrL}.
\item[2.] Factor $c_L$ using Mathematica's {\tt Factor} command \cite{Wolfram_88}.
\item[3.] Apply the map $\mathscr{L}^{-1}$ to each factor in $R[L]$ from the previous step.
In particular, ${\mathscr L}^{-1}(l_{i_1}l_{i_2}\cdots l_{i_k})=l_{i_1}\shuffle l_{i_2}\shuffle\cdots\shuffle l_{i_k}$.
\end{customlist}
An efficient algorithm for the Chen-Fox-Lyndon factorizations in step 1 is given in \cite{Duval_83}
(see also \cite{Scholtens_94}).
Mathematica's implementation notes for {\tt Factor} provide a description of the
specific algorithms used to factor multivariate polynomials.
For a more general treatment of the subject see \cite{von_zur_Gathen_85}.

\begex
{\rm
Consider the polynomial
\begin{align*}
c&=
2 x_0^2 - 2 x_1^2 + \underline{2 x_0^2x_1x_0 + 2 x_0x_1x_0^2} - 2 x_0x_1^2x_0 \\
&\hspace*{0.16in}+ 2 x_1x_0^2x_1 + 2 x_1x_0x_1^2 +2 x_1^2x_0x_1 + 2 x_0x_1x_0x_1x_0x_1 \\
&\hspace*{0.16in}+ 2 x_0x_1x_0x_1^2x_0 +4 x_0x_1^2x_0^2x_1 + 2 x_0x_1^2x_0x_1x_0 \\
&\hspace*{0.16in}+ 2 x_1x_0^2x_1x_0x_1 + 4 x_1x_0^2x_1^2x_0 + 2 x_1x_0x_1x_0^2x_1 \\
&\hspace*{0.16in}+ 2 x_1x_0x_1x_0x_1x_0,
\end{align*}
which does not have relative degree since it has no linear words of the form
$x_0^{r-1}x_1$ in its support.
The algorithm above is applied to $c$
with the help of the Mathematica NCFPS package \cite{NCFPS_21}.
The underlined terms correspond to Example~\ref{ex:mathscrL} which is embedded in this example.

\noindent\underline{Step 1}: Observe
\begin{align*}
c_L&={\mathscr L(c)}=l_0^2 - l_1^2 + \underline{l_0^2 l_2} + l_1^2 l_2 + l_0 l_1 l_2^2 -\underline{2 l_0 l_3} \\
&\hspace*{0.15in}+ 2 l_1 l_3 - 2 l_1 l_2 l_3 - 2 l_0 l_4 - 2 l_1 l_4 - 2 l_0 l_2 l_4 + 4 l_3 l_4.
\end{align*}

\noindent\underline{Step 2}: Using the $\tt Factor$ command in Mathematica gives
\begdi
c_L=(l_0 + l_1 + \underline{l_0 l_2 - 2 l_3}) (\underline{l_0} - l_1 + l_1 l_2 - 2 l_4).
\enddi

\noindent\underline{Step 3}: Mapping each factor of $c_L$ back to $\allpolyprop$ yields
\begdi
c=c_1\shuffle c_2=(x_0 + x_1 + \underline{x_0x_1x_0})\shuffle (\underline{x_0} - x_1 + x_1x_0x_1).
\enddi

Observe that the two factors of $c$ are distinct and linearly nullable with relative degree $r=1$.
Hence, there exist two distinct
nulling inputs $c_{u^\ast_1}$ and $c_{u^\ast_2}$ for this polynomial.
Each input can be computed via the algorithm in \cite{Gray-etal_Automatic_14}
or by solving an initial value problem
which follows from setting $F_{c_i}[u]=0$ and then repeatedly differentiating with respect to time.
For $c_1$ the latter approach yields
\begdi
u\ddot u-2\dot u^2-u^4=0,\;\;u(0)=-1,\;\;\dot u(0)=0
\enddi
so that
\begdi
c_{u^\ast_1} = \mbf{1} + x_0^2 + 7 x_0^4 +127 x_0^6 + 4369 x_0^8+\cdots.
\enddi
Similarly, for $c_2$
the corresponding initial value problem is
\begdi
\dot u+tu=0,\;\; u(0)=-1,
\enddi
which gives
\begdi
c_{u^\ast_2}=
-\mbf{1} + x_0^2 - 3 x_0^4 +
15 x_0^6 - 105 x_0^8+\cdots.
\enddi

To empirically verify that $c\circ c_{u^\ast_i}=0$, it is necessary to truncate $c_{u^\ast_i}$.
This means that $c\circ c_{u^\ast_i}$ will not be exactly
zero, but instead zero up to some word length depending on the number of terms retained in
$c_{u^\ast_i}$. For example,
truncating both $c_{u^\ast_i}$
to words of maximum length six gives
\begin{align*}
c\circ c_{u^\ast_1}&=87 380 x_0^{10} +
2 946 560 x_0^{12} + 153 856 528 x_0^{14}+O(x_0^{16}) \\
c\circ c_{u^\ast_2}&=2100 x_0^{10} -840 840 x_0^{14} +
57 657 600 x_0^{16} - O(x_0^{18}).
\end{align*}
}
\endex

\begex \label{ex:lnull-shuffle-non-null-v2}
{\rm
Reconsider Example~\ref{ex:lnull-shuffle-non-null} where $c_L=l_0+l_1$ and $d_L=\mbf{1}+l_1$.
As observed earlier, $c\shuffle d$ is primely nullable and linearly nullable. Clearly
$(c\shuffle d)_L=c_Ld_L$ is reducible with one linearly nullable factor $c_L$.
}
\endex

\begex
{\rm
Recall that for polynomials in one variable, the class of irreducible polynomials depends on the base field. For example,
over the real field, the irreducible polynomials are either of degree $1$ or degree $2$ (e.g., $x_0^2+1$). Over the complex field,
there are only degree $1$ irreducibles \cite[Chapter IV.1]{Lang02}. However, in every multivariate polynomial ring there are irreducible
elements of higher degree.
Consider the polynomial $c=6x_1^3-2x_1x_0^2-2x_0x_1x_0-2x_0^2x_1-24x_0^4\in\allpolyprop$.
It does not have relative degree, and thus, it is not linearly nullable. There is at present no direct test for any other form of nullability.
In the Lyndon basis,
it follows that $c_L = l_1^3-l_0^2l_1-l_0^4\in\re[L]$. Now if $c_L$ is reducible, one could write
\begeq \label{eq:c_L-irreducible-factorization}
c_L=(l_1-p_1(l_0)) (l_1^2+p_2(l_0)l_1+p_3(l_0))
\endeq
for some polynomials $p_i(l_0)$. Since $l_0^4=p_1(l_0)p_3(l_0)$, necessarily $p_1(l_0)=al_0^n$ and $p_3(l_0)=bl_0^{4-n}$ for some $n \in \{0,1,2,3,4\}$ and $a,b \in \re$ with $ab=1$.
Substituting these forms into \rref{eq:c_L-irreducible-factorization} shows directly that there are no values of $n$ that can yield $c_L$.
Thus, $c_L$ is an irreducible multivariate polynomial of degree $4$ as an element in $\re[L]$.
}

\endex

\section{Conclusions}

Working entirely in a Chen-Fliess series setting, it was shown that
the problem of zeroing the output can be solved for the class of systems
who generating series can be factored via the shuffle algebra into
terms where at least one factor is nullable.
The shuffle algebra on $\allpoly$ was shown to be a unique factorization domain so that
any polynomial can be uniquely factored into its irreducible elements for this purpose.
The factorization is done by viewing this
shuffle algebra as the symmetric algebra over the vector space spanned by Lyndon words.
A specific polynomial factorization algorithm was given based on the Chen-Fox-Lyndon factorization of words.

\section*{Acknowledgments}

KEF and AS were supported by the Research Council of Norway through project 302831 Computational Dynamics and Stochastics on Manifolds (CODYSMA).

The authors wish to thank Lance Berlin for maintaining the Mathematica software package
Noncommutative Formal Power Series (NCFPS),
which was used extensively in this paper to do the formal power series
calculations.

\begin{appendix}
\vspace*{0.15in}
\noindent
\vspace*{0.1in}
{\bf Appendix. Proof of Identity \rref{eq:left-shift-composition-product}}

Observe for any $c,d\in\allseries$ that
\begin{align*}
x_0^{-1}(c\circ d)&= \sum_{\eta\in X^{\ast}} (c,\eta)\,x_0^{-1}(\eta\circ d)\\
&= \sum_{\eta\in X^{\ast}} (c,x_0\eta)\,x_0^{-1}(x_0\eta\circ d) + \\
&  \hspace*{0.25in}\sum_{i=1}^m\sum_{\eta\in X^{\ast}}
(c,x_i\eta)\, x_0^{-1}(x_i\eta\circ d) \\
&= \sum_{\eta\in X^{\ast}} (x_0^{-1}(c),\eta)\,\eta\circ d + \\
&  \hspace*{0.2in} \sum_{i=1}^m\sum_{\eta\in\eta^{\ast}}(x_i^{-1}(c),\eta)\, (d_i\shuffle(\eta\circ d)) \\
&= x_0^{-1}(c)\circ d + \sum_{i=1}^m d_i\shuffle  [x_i^{-1}(c)\circ d].
\end{align*}
For any $i\neq 0$, note that
\begdi
x_i^{-1}(c\circ d)= \sum_{\eta\in X^\ast} (c,\eta)\, x_i^{-1}(\eta\circ d),
\enddi
and $\eta\circ d$ always has a leading $x_0$ when $\eta$ is nonempty. Thus,
$x_i^{-1}(\eta\circ d)=0$ for all $\eta\in X^\ast$, and the identity is verified.
\end{appendix}

\end{document}